\documentclass[pdflatex,sn-nature]{springernature/sn-jnl}

\usepackage{graphicx}%
\usepackage{multirow}%
\usepackage{amsmath,amssymb,amsfonts}%
\usepackage{amsthm}%
\usepackage{mathrsfs}%
\usepackage[title]{appendix}%
\usepackage{xcolor}%
\usepackage{textcomp}%
\usepackage{manyfoot}%
\usepackage{booktabs}%
\usepackage{algorithm}%
\usepackage{algorithmicx}%
\usepackage{algpseudocode}%
\usepackage{listings}%
\usepackage{siunitx}

\theoremstyle{thmstyleone}%
\theoremstyle{thmstyletwo}%

\theoremstyle{thmstylethree}%

\unnumbered

\usepackage{etoolbox}
\usepackage{geometry}
\usepackage{microtype}
\usepackage{ragged2e}
\usepackage{caption}
\usepackage{pdflscape}
\usepackage{adjustbox}
\usepackage{xparse}
\usepackage{tabularx}
\usepackage{tikz}
\usepackage[section]{placeins}

\renewcommand{\baselinestretch}{1.15}
\makeatletter
\def\ps@plain{%
  \let\@mkboth\@gobbletwo
  \let\@oddhead\@empty
  \let\@evenhead\@empty
  \def\@oddfoot{\hfil{\normalfont\fontsize{9}{11}\selectfont\thepage}\hfil}%
  \let\@evenfoot\@oddfoot
}
\makeatother

\AtBeginDocument{%
  \pagestyle{plain}%
  \raggedbottom
  \justifying
  \setlength{\parindent}{1.5em}%
  \setlength{\parskip}{0pt plus 1pt}%
  \ifcsdef{bibhang}{\setlength{\bibhang}{1.5em}}{}%
  \ifcsdef{bibsep}{\setlength{\bibsep}{3pt plus 1pt}}{}%
}

\newcommand{\RevisedTableAlias}{}
\newcommand{\LabelRevisedTable}{%
  \ifdefempty{\RevisedTableAlias}{}{\label{\RevisedTableAlias}}%
}
\AtEndEnvironment{longtable}{\LabelRevisedTable}
\newlength{\ManuscriptTableMaxHeight}

\newcommand{\ManuscriptFitTable}[1]{%
  \par\noindent
  \begin{adjustbox}{max totalsize={\linewidth}{\ManuscriptTableMaxHeight},center}
    \begin{minipage}{\linewidth}
      \captionsetup{type=table}%
      \renewcommand{\baselinestretch}{1}\small
      \setlength{\parindent}{0pt}%
      \setlength{\parskip}{0pt}%
      \centering
      #1\par
      \LabelRevisedTable
    \end{minipage}
  \end{adjustbox}\par
}

\newcommand{\ManuscriptPrepareTables}{%
  \RenewDocumentEnvironment{table}{O{htbp} +b}{\ManuscriptFitTable{##2}}{}%
  \RenewDocumentEnvironment{table*}{O{htbp} +b}{\ManuscriptFitTable{##2}}{}%
  \ifcsdef{sidewaystable}{%
    \RenewDocumentEnvironment{sidewaystable}{O{htbp} +b}{\ManuscriptFitTable{##2}}{}%
  }{}%
  \ifcsdef{sidewaystable*}{%
    \RenewDocumentEnvironment{sidewaystable*}{O{htbp} +b}{\ManuscriptFitTable{##2}}{}%
  }{}%
}

\newif\ifManuscriptLandscapePage
\AddToHook{shipout/foreground}[manuscript-landscape-folio]{%
  \ifManuscriptLandscapePage
    \begin{tikzpicture}[remember picture,overlay]
      \node[rotate=90,anchor=base east,inner sep=0pt,outer sep=0pt]
        at ([xshift=-13mm,yshift=-25mm]current page.north east)
        {\normalfont\fontsize{9}{11}\selectfont\thepage};
    \end{tikzpicture}%
  \fi
}

\newenvironment{ManuscriptLandscapeTable}{%
  \begin{landscape}
  \ManuscriptLandscapePagetrue
  \pagestyle{empty}%
  \setlength{\textwidth}{\linewidth}%
  \setlength{\columnwidth}{\linewidth}%
  \setlength{\ManuscriptTableMaxHeight}{\dimexpr\textheight-2\baselineskip\relax}%
}{%
  \end{landscape}
}

\begin{document}
\begin{center}
\begin{minipage}{0.96\textwidth}
\centering
\setlength{\parskip}{0pt}
{\fontsize{18}{22}\selectfont\bfseries
\textbf{Project financing in vanishing forests:}\\[3pt]
\textbf{Do investors care about nature protection?}
\par}
\vspace{12pt}
{\normalsize
\mbox{Theodor Cojoianu\textsuperscript{1,2,3}},\quad
\mbox{Mohammad Faour\textsuperscript{4}},\quad
\mbox{Andreas Hoepner\textsuperscript{3,5}}\\[4pt]
\mbox{Matthew Purslow\textsuperscript{6}},\quad
\mbox{Khaladdin Rzayev\textsuperscript{6,7}},\quad
\mbox{Ruoran Zhao\textsuperscript{8}}\par}
\vspace{10pt}
{\fontsize{9}{12}\selectfont
\textsuperscript{1}College of Integrative Studies, Singapore Management University, Singapore\par
\textsuperscript{2}Bucharest University of Economic Studies, Bucharest, Romania\par
\textsuperscript{3}European Commission Platform on Sustainable Finance, Brussels, Belgium\par
\textsuperscript{4}Suliman S. Olayan School of Business, American University of Beirut, Beirut, Lebanon\par
\textsuperscript{5}University College Dublin, Dublin, Ireland\par
\textsuperscript{6}Business School, The University of Edinburgh, Edinburgh, United Kingdom\par
\textsuperscript{7}Koç University, Istanbul, Turkey\par
\textsuperscript{8}University of York, York, United Kingdom\par}
\vspace{8pt}
{\footnotesize Correspondence: \href{mailto:tcojoianu@smu.edu.sg}{tcojoianu@smu.edu.sg}\par}
\end{minipage}
\end{center}
\vspace{2pt}
\section*{Abstract}
\noindent
Biodiversity risks are increasingly recognized by financial market participants, yet their measurement at the corporate-entity level obscures the locations and assets through which they arise. Project finance links identifiable physical assets to dedicated financing contracts, allowing us to examine this transmission more directly. By linking 9,037 project–loan pairs in 115 countries, we relate bank loan spreads to projects’ proximity to protected areas, surrounding forest loss and lenders’ nature policies. We find that lenders price proximity to protected areas only when they also have explicit nature risk policies and particularly within 1.5 km buffers from the project location. Realised forest loss around a project is priced more consistently than forest loss after the loan is signed, and risk pricing near protected areas strengthens after the Kunming–Montreal Global Biodiversity Framework. Listed company sponsors also experience negative, but short-lived, abnormal returns around loan signings for projects within 30 km of protected areas or in high-forest-loss landscapes. Our results suggest that while nature risk is partially priced in project finance, this is not a strong incentive to deter significant harm to nature.
\newpage
\noindent
Around one-third of the world's protected land is under intense human pressure, including from roads, mining, energy facilities, and other infrastructure \cite{jones_one-third_2018, golden_kroner_uncertain_2019, rehbein_renewable_2020}. The financial sector is increasingly recognized as a key actor who can disincentivise biodiversity loss, given its role in funding agriculture and infrastructure projects that place additional pressure on nature. The adoption of the Kunming–Montreal Global Biodiversity Framework (GBF) in December 2022 brought nature to the forefront of sustainable finance policies. GBF's Targets 14 and 15 call for financial flows to be progressively aligned with the framework and for large companies and financial institutions to assess and disclose their biodiversity risks, dependencies and impacts \cite{cbd_kunming-montreal_2022}. Central banks and supervisors now regard nature degradation as a potential threat to the stability of the financial system \cite{ngfs_nature_2024}. For example, in the euro area, around 72\% of companies are highly dependent on at least one ecosystem service, and around three-quarters of corporate loans go to such firms \cite{svartzman_silent_spring_2021,ceglar_european_2025}. Whether banks price biodiversity risks is therefore an important question, both for halting biodiversity loss and for financial stability. 

Emerging evidence also suggests that, beyond central banks,  companies and financial market participants also consider nature-related risks as financially material \cite{carvalho_impacts_2023,gjerde_corporate_2026, giglio_biodiversity_2026}. Nature factors are relevant for asset prices, particularly through the risk of future policy inspired from the GBF and the adoption of the Taskforce for Nature-related Financial Disclosures  (TNFD) \cite{garel_investors_2024}. Recent evidence shows that biodiversity exposure is linked to equity values \cite{garel_investors_2024,garel_firm-level_2026}, corporate bond spreads \cite{cherief_novel_2025}, bank loan pricing \cite{becker_loan_2025, li_banks_2025, hirschbuhl_climate_2025, bax_role_2026} and sovereign creditworthiness \cite{agarwala_biodiversity_2026}. However, scientifically quantifying nature-related risks has been challenging. A listed company or corporate borrower is exposed to nature risks across numerous sites, geographies, industries and asset types, and its supply chains connect it to distant ecosystems. Equity investors, for instance, penalise firms' direct biodiversity impacts but largely ignore those embedded in supply chains, which account for most biodiversity loss \cite{duan_when_2026}. Given that there is no biodiversity metric analogous to the ton of CO$_2$-equivalent with which such exposures could be added and compared to across geography and time \cite{kedward_biodiversity_2023, wauchope_what_2024}, these studies  rely mainly on indirect proxies, such as sector-average footprints, text-based measures and environmental, social and governance (ESG) ratings \cite{garel_investors_2024, giglio_biodiversity_2026, xin_noisy_2025}. Such proxies merely infer exposure rather than measure it directly.

Since  nature-related risk exposures are inherently location-specific \cite{ngfs_nature_2024}, they would ideally be measured where they arise; at the location of each physical asset and relative to the ecosystems around it. This is consistent with the TNFD disclosure framework, which recommends that organizations report the locations of their assets and activities in ecologically sensitive areas \cite{tnfd_recommendations_2023}. So far, studies of housing and land show how local ecosystems influence property values \cite{liu_financial_2026,manning_ecosystem_services_2022}, while municipal bond studies connect ecosystem degradation and biodiversity protection to public borrowing costs \cite{chen_pricing_2024,rizzi_natural_capital_2022}. Research examining industrial facilities further links the designation of nearby protected areas to the facilities’ output and to the market value of their parent firms \cite{akbari_real_2026}. These settings highlight the financial materiality of an asset's location relative to nature. The question for bank lending is therefore whether an asset’s proximity to ecologically sensitive areas is priced in the loans that fund it, and if this relationship depends on the lenders' nature commitments.

Project finance provides a direct setting in which we can examine this question. The asset class underpins large, long-lived investments across energy, utilities, transport and other infrastructure industries, where project viability and loan risk depends on the location and operation of the underlying physical asset \cite{esty_why_2004}. Project sponsors establish and invest equity in a legally separate purpose vehicle (SPV)  that holds the project's assets. The project company borrows in its own name, servicing its debt primarily from the project’s cash flows \cite{byoun_risk_2013}. Lenders therefore have limited or no recourse to the sponsors, with their claims limited to the project assets and contractual rights \cite{corielli_risk_2010}. This structure links identifiable assets at known locations to dedicated financing contracts and the banks arranging them, allowing us to examine how local nature exposure is incorporated into loan pricing. Unlike corporate-level measures that combine exposure across numerous operations, project-level measures retain the geographical specificity of that exposure. Although project debt and equity are predominantly privately negotiated, loan records reveal financing terms, while listed sponsors’ share prices provide a complementary view, unveiling listed equity investor responses when loans are signed \cite{kammoun_capital_2021}.

\begin{figure}[!htbp]
\centering
\includegraphics[width=0.96\linewidth]{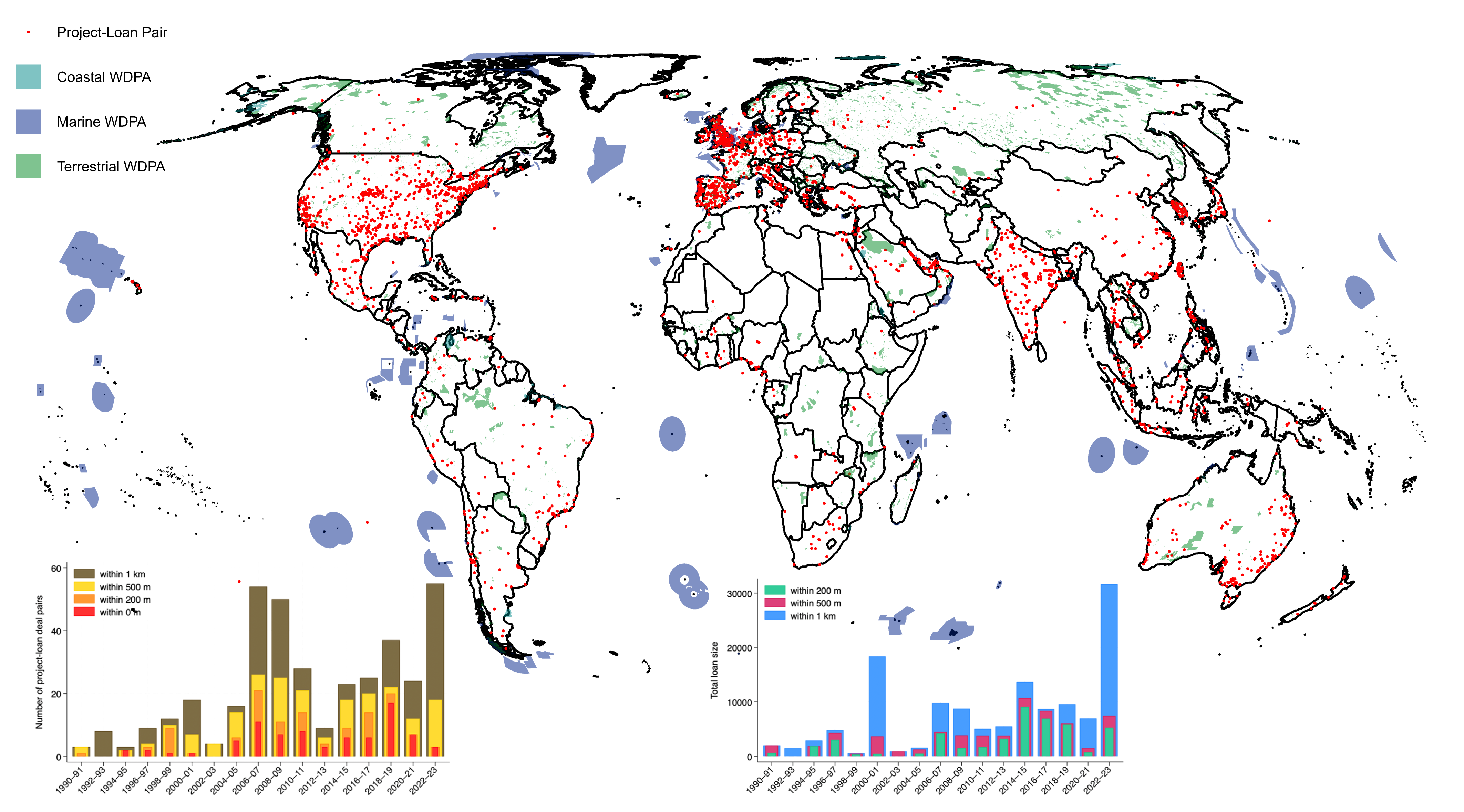}
\caption{\textbf{Project finance and protected areas.} Global distribution of project--loan pairs, with inset panels showing the number of pairs and total loan size over time, grouped by proximity to entries in the World Database on Protected Areas (WDPA). The overall sample comprises 9,037 pairs associated with 3,793 projects in 115 countries. Points represent project locations and can overlap.}
\label{fig:map}
\end{figure}

Thus, we connect 3,793 projects and 9,037 project--loan pairs across 115 countries (Fig.~\ref{fig:map}) with protected-area boundaries, satellite-derived forest loss and arranging banks’ nature policies. Our analysis makes four contributions. First, we estimate how loan spreads vary with each project's proximity to protected areas and with the already realised forest loss in the surrounding landscape, extending evidence on the financial materiality of nature to project financing. Second, we examine whether this relationship varies with the lenders’ own nature-related policies. Third, by measuring proximity to protected areas and forest loss across multiple buffers, we assess whether pricing is concentrated near the legal boundaries of protected areas or extends across the wider landscape. Finally, we distinguish forest loss observed before financing from loss realised afterwards to assess whether loan terms anticipate subsequent forest loss or merely reflect existing damage. We also examine whether pricing changed after the adoption of the GBF, and how sponsors’ share prices respond when loans are signed. We find that banks price nature exposure near protected areas only when they hold nature-related policies and that loan terms also reflect forest loss that has already occurred, but only weakly anticipate subsequent loss.

\newpage
\section*{Results}\label{sec:results}

\subsection*{Lenders' nature policies shape loan pricing around protected areas}

Our global project--loan pairs dataset spans all inhabited continents, with the densest activity in North America, Europe and South and Southeast Asia (Fig.~\ref{fig:map}). Oil and gas (2,692 pairs) and utilities (2,232 pairs) account for more than half of the sample (Supplementary Table~\ref{revtab:industry}). We find that proximity to protected areas is common, as the median project--loan pair lies 40.5~km from the nearest protected area designated before loan signing, and a quarter lie within 7.4~km (Supplementary Table~\ref{revtab:summary_statistics}). In total, 862 project--loan pairs (9.5\%) are located within 2~km of a protected area, including 158 inside protected area boundaries.

\begin{figure}[!htbp]
    \centering
    \includegraphics[width=\textwidth]{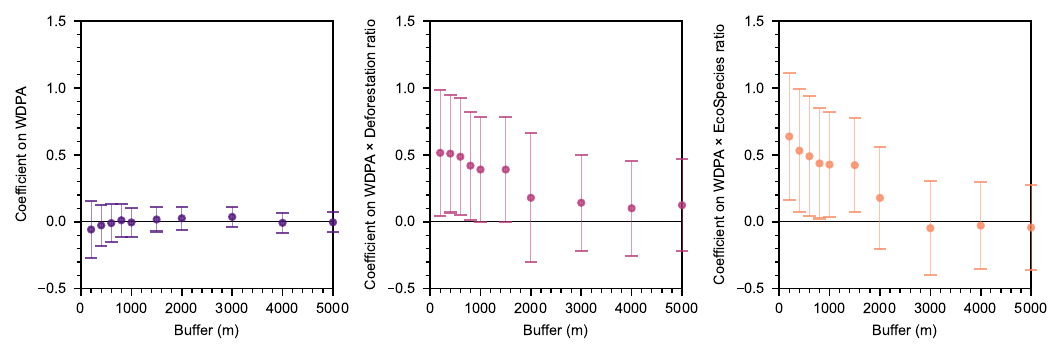}
    \caption{\textbf{Lender nature policies condition the pricing of protected-area proximity.} Coefficients (points) and uncertainty intervals (error bars) from regressions of the log loan spread on an indicator for projects located within a given distance (horizontal axis) of the nearest protected area, estimated separately for each distance. Left, coefficient on the proximity indicator (from the EcoSpecies models). Centre, coefficient on the interaction between proximity and the Deforestation ratio (share of bookrunners with deforestation-related policies). Right, coefficient on the interaction between proximity and the EcoSpecies ratio (share of bookrunners with ecosystem- or species-related policies). Policy ratios are on a 0--1 scale. Interaction coefficients describe how the proximity--spread relationship changes with policy coverage. All models include the loan and project controls and the fixed effects described in Methods; standard errors are clustered by project; $n=6{,}907$ project--loan pairs. Full estimates, including those for projects inside protected areas, are reported in Supplementary Tables~\ref{tab:EcoSpecies_wdpa} and~\ref{tab:deforestation_wdpa}.}
    \label{fig:coefficient_WDPA}
\end{figure}

We first examine the relationship between loan spreads and project proximity to protected areas, and in particular we test whether it varies with the share of banks coordinating the loan (i.e. bookrunners) in the loan syndicate that have  nature-related policies (see \hyperref[sec:methods]{Methods}). When no bookrunner holds an ecosystem-related or species-related policy, proximity to protected areas is not significantly associated with loan spreads at any distance threshold ($p\geq0.10$ throughout; Fig.~\ref{fig:coefficient_WDPA}, left; Supplementary Table~\ref{tab:EcoSpecies_wdpa}). 

By contrast, proximity to protected areas is associated with higher loan spreads for syndicates with a higher share of bookrunners holding ecosystem-related or species-related policies. The interaction between proximity and the EcoSpecies ratio  -- the share of bookrunners with ecosystem-related or species-related policies -- is positive and statistically significant at thresholds from 200~m to 1.5~km. The coefficients decline as the distance threshold increases, from $\beta=0.637$ ($p<0.01$) at 200~m to $\beta=0.423$ ($p<0.05$) at 1.5~km, indicating that the pricing difference is strongest for projects closest to protected areas (Fig.~\ref{fig:coefficient_WDPA}, right). To illustrate the economic magnitude, our estimates suggest that at the 200~m threshold, a 10 percentage point increase in the share of bookrunners with nature-related policies is associated with a 6.6\% ($ \approx e^{0.637 \times 0.10}-1$) increase in loan spreads, holding other covariates constant. The interaction terms are statistically insignificant at thresholds of 2~km and beyond. Inside protected areas, the interaction is positive but statistically insignificant ($\beta=0.521$,  $p\geq0.10$), which may reflect the small number of observations for projects located inside protected areas (158 project-loans). The results are similar when the EcoSpecies ratio is replaced with a binary EcoSpecies policy indicator. The interaction coefficients decline from $\beta=0.429$ ($p<0.05$) inside protected areas to $\beta=0.172$ ($p<0.05$) at 1.5~km, becoming statistically insignificant at 2~km ($\beta=0.082$, $p\geq0.10$; Supplementary Table~\ref{revtab:EcoSpecies_binary}).

Next, we examine whether the relationship between proximity and loan spreads also varies with the bookrunners' deforestation policies. As with the EcoSpecies ratio, the interaction term between the deforestation ratio and proximity is positive and statistically significant at distance thresholds from 200~m to 1.5~km. The coefficients range from $\beta=0.390$ to $\beta=0.515$ ($p<0.05$ throughout), falling to $\beta=0.179$ ($p\geq0.10$) at 2~km (Fig.~\ref{fig:coefficient_WDPA}, centre; Supplementary Table~\ref{tab:deforestation_wdpa}). Overall, these findings suggest that proximity to protected areas is priced into loan spreads when bookrunners hold nature-related policies.

\subsection*{Forest loss and loan pricing}
Protected areas capture the legal status of nature around a project, whereas forest loss is a more narrow, dynamic indicator that captures one aspect of change in the surrounding landscape.  To understand whether forest loss is priced into project loans., we  measure the cumulative non-burn forest loss over the available years up to ten years before loan signing. We interact this measure with the share of bookrunners holding deforestation policies (\hyperref[sec:methods]{Methods}). 

\begin{figure}[!htbp]
    \centering
    \includegraphics[width=\textwidth]{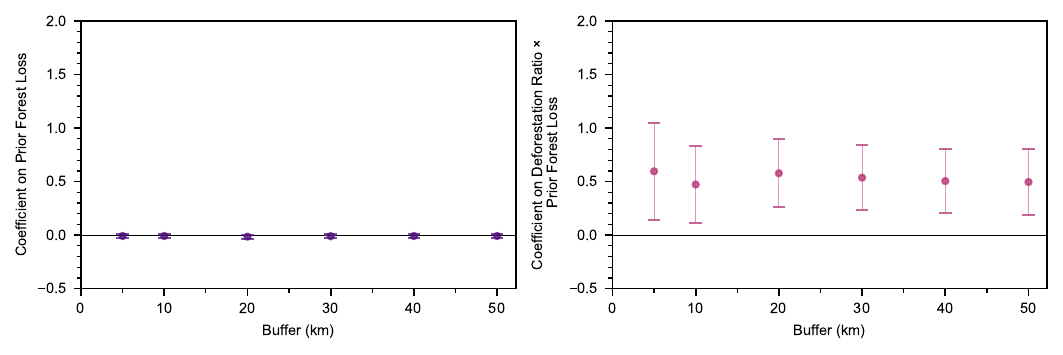}
    \caption{\textbf{Lenders with deforestation policies price forest loss that precedes loan signing.} Coefficients (points) and uncertainty intervals (error bars) from regressions of the natural logarithm of the loan spread on $\ln(1+F)$, where $F$ is cumulative non-burn forest loss in the available years up to ten years before loan signing, measured within buffers of 5--50~km around each project (horizontal axis). Left, coefficient on $\ln(1+F)$. Right, coefficient on its interaction with the Deforestation ratio, which ranges on a scale from 0 to 1. All models include the controls and fixed effects described in Methods; standard errors are clustered by project; $n=721$--$728$ project--loan pairs. Estimates for forest loss before and after signing are reported in Supplementary Tables~\ref{revtab:prior_forest_loss} and~\ref{revtab:subsequent_forest_loss}; the latter are not displayed.}
    \label{fig:coefficient_Forest_Loss}
\end{figure}

When no bookrunner holds a deforestation policy, prior forest loss is not significantly associated with loan spreads at any radius between 5~km and 50~km ($\beta=-0.014$ to $-0.008$, $p\geq0.10$ throughout; Fig.~3, left; Supplementary Table~9). By contrast, greater prior forest loss is associated with higher loan spreads as the share of bookrunners holding deforestation policies increases. The interaction term between prior forest loss and the Deforestation ratio is positive and statistically significant across all radii examined, ranging from $\beta=0.473$ to $\beta=0.596$ ($p<0.01$ throughout; Fig.~\ref{fig:coefficient_Forest_Loss}, right). Hence, we unveil that spreads rise more steeply with the amount of forest loss that a landscape has already experienced, but only when more of the arranging banks pay attention to deforestation through their policies.

We then examine whether lenders anticipate expected future forest losses in the pricing of project loans by relating loan spreads to forest losses realised up to ten years following the loan. The results tell a less consistent story, but provide some evidence that lenders with deforestation policies price expected future forest loss. The interactions between subsequent forest loss and the Deforestation ratio are positive across all radii, although statistically weaker, ranging from $\beta=0.121$ ($p\geq0.10$) to $\beta=0.156$ ($p<0.10$) (Supplementary Table~\ref{revtab:subsequent_forest_loss}). These models use a larger sample (4,677 pairs) and different observation windows, so the two sets of magnitudes are not directly comparable. The evidence is therefore more consistent for the pricing of previously observed forest loss than for the anticipation of subsequent realised loss.

\subsection*{Proximity pricing after the Global Biodiversity Framework}
\begin{figure}[!htbp]
    \centering
    \includegraphics[width=0.78\textwidth]{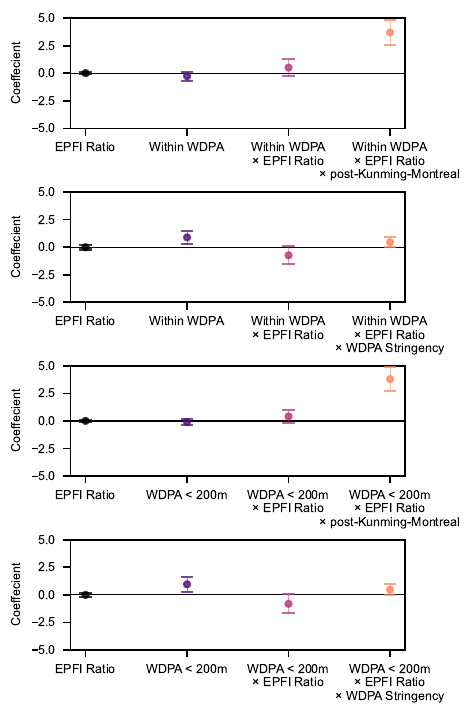}
    \caption{\textbf{The pricing of protected-area proximity appears stronger after the Kunming--Montreal Global Biodiversity Framework.} Coefficients (points) and uncertainty intervals (error bars) on the EPFI ratio (share of bookrunners that are Equator Principles Financial Institutions), protected-area proximity, their interaction and their triple interaction with either an indicator for loans signed after 19 December 2022 (first and third panels) or the management category of the nearest protected area on a reversed IUCN scale from 1 to 6 (second and fourth panels), from regressions of the log loan spread. Proximity is defined as location inside a protected area (first and second panels) or within 200~m of one (third and fourth panels). The EPFI ratio is on a 0--1 scale. All models include the controls and fixed effects described in Methods; standard errors are clustered by project; $n=6{,}907$ (post-framework models) and $n=3{,}720$ (category models). Full estimates are reported in Supplementary Table~\ref{revtab:EPFIs_wdpa_kunming}, and corresponding estimates for the EcoSpecies ratio in Supplementary Table~\ref{tab:EcoSpecies_wdpa_kunming}.}
    \label{fig:coefficient_EPFIs}
\end{figure}

In this part of the analysis, we interact proximity and lender policy coverage with an indicator for loans signed after the adoption of the GBF on 19 December 2022 (see \hyperref[sec:methods]{Methods}). For projects inside and within 200~m of a protected area, respectively, we observe a significant increase in their loan spreads with $\beta=6.407$ ($p<0.01$) and $\beta=6.197$ ($p<0.01$) for the EcoSpecies ratio (Supplementary Table~\ref{tab:EcoSpecies_wdpa_kunming}), and $\beta=3.701$ ($p<0.01$) and $\beta=3.798$ ($p<0.01$) for the share of bookrunners that are Equator Principles Financial Institutions (EPFI ratio; Fig.~\ref{fig:coefficient_EPFIs}; Supplementary Table~\ref{revtab:EPFIs_wdpa_kunming}). 

Pricing also varies with the nearest protected area's management objectives, with proximity to more stringent protected areas signifying greater financial risk for projects of EPFI signatories. The triple interactions with this score are $\beta=0.465$ ($p<0.05$) inside protected areas and $\beta=0.471$ ($p<0.05$) within 200~m for the EcoSpecies ratio, and $\beta=0.455$ ($p<0.10$) and $\beta=0.468$ ($p<0.10$), respectively, for the EPFI ratio. Because IUCN categories describe management objectives rather than enforcement~\cite{dudley_guidelines_2008}, we read this as heterogeneity across types of protection rather than as a response to enforcement itself.

\subsection*{Sponsor equity responses to project nature risk exposure}
\begin{figure}[!htbp]
    \centering
    \includegraphics[width=0.92\textwidth]{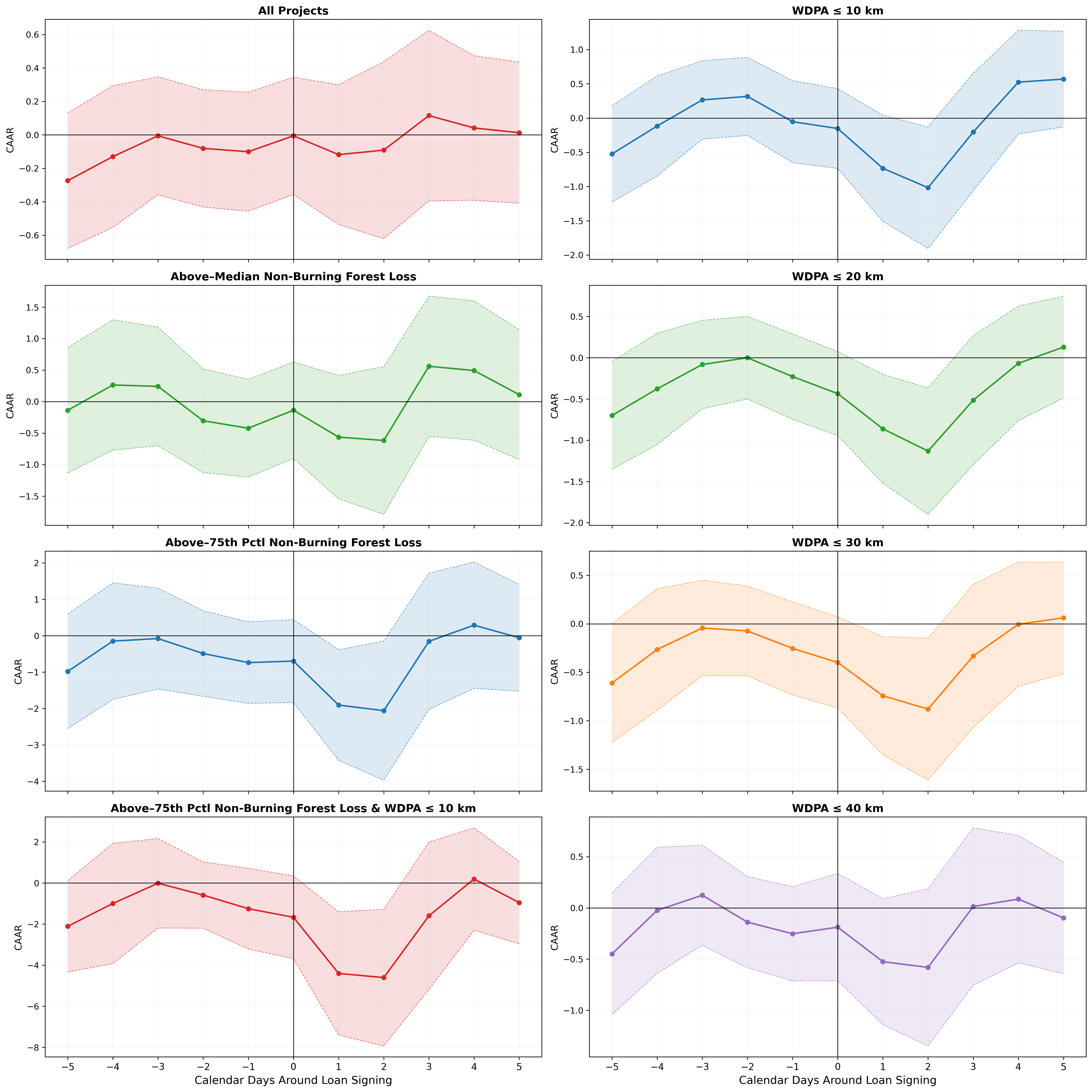}
    \caption{\textbf{Sponsor share prices around the signing of project loans with high nature exposure.} Cumulative average abnormal returns (CAARs, in per cent; solid lines) and 95\% confidence intervals (shaded areas bounded by dashed lines) for project sponsors over calendar days $-5$ to $+5$ around the signing of a project loan. CAARs are cumulated from day $-15$. Left column, from top: all signings ($n=1{,}900$); projects with above-median non-burn forest loss; projects with above-75th-percentile non-burn forest loss; and projects with above-75th-percentile non-burn forest loss located within 10~km of a protected area. Right column, from top: projects located within 10, 20, 30 and 40~km of a protected area. Sub-samples overlap. Expected returns are estimated with the developed-market five-factor model (Methods).}
    \label{fig:event_study_combined}
\end{figure}

Fig.~\ref{fig:event_study_combined} plots cumulative average abnormal returns (CAARs) around 1,900 project loan signings by 601 publicly listed sponsors headquartered in developed markets. For the full sample, the CAAR is not statistically different from zero over days $-5$ to $+5$, consistent with earlier evidence that sponsors do not gain from project loan announcements~\cite{kammoun_capital_2021}. However, the results differ for projects with exposure to nature risk. For projects located within 10, 20 or 30 km of a protected area, CAARs are negative two days after loan signing, and the 95\% confidence intervals do not include zero. For projects within 40 km of a protected area, the decline is smaller and the confidence intervals include zero. Projects in the top quartile of non-burn forest loss in the ten years after signing have negative CAARs on the day after signing, and the decline is larger for projects that are also within 10 km of a protected area. For projects with above-median forest loss, CAARs are not statistically different from zero. These negative reactions are short-lived, and CAARs return towards zero within four days. We also run additional models for burnt loss surrounding the project before the loan announcement, but we find no statistically significant results, which suggests perhaps that equity markets are much more concerned with the forward-looking forest loss embedded risk, then realized forest loss around the location of an announced project. 

\section*{Discussion}\label{sec:discussion}

\subsection*{Pricing of nature risk in project finance}
Our results show that certain elements of nature risk are priced in project finance. The first important element in our paper concerns lenders: banks with nature-related policies charge higher loan spreads on projects close to protected areas and in landscapes with greater prior forest loss. This is compatible with the lower average spreads paid by borrowers of Equator Principles banks~\cite{amiram_economic_2021,power_does_2025}, since a lender can offer cheaper finance overall while distinguishing between exposed and less exposed projects to ecologically sensitive areas. Pricing, however, is neither credit rationing nor environmental performance. Equator Principles members have not financed fewer hazardous projects than non-members~\cite{eshet_sustainable_2017}, and we observe only the projects that were financed. We cannot tell whether committed lenders also deny credit to the most exposed projects, or whether those projects are simply financed by other lenders.

The second important element in our paper concerns space. Lenders with nature-related policies price proximity to protected areas only up to a 1.5~km buffer zone. The ecological footprint of infrastructure projects, however, extends much further. For instance, mammal populations decline up to about 5~km from infrastructure~\cite{benitezlopez_impacts_2010}, and mining in the Brazilian Amazon has been linked to forest loss up to 70~km beyond lease boundaries~\cite{sonter_mining_2017}, also due to the indirect effects of mining rather than the mining lease itself. One potential explanation is that lenders respond to the legal boundary of a protected area itself, rather than to how far a project's ecological impact extends. The boundary is where project permits become harder to obtain and projects attract stricter due diligence \cite{ifc_ps6_2012}. This points to a potential underpricing of nature risk beyond protected area boundaries. Pricing should therefore complement, rather than substitute for, ecological assessments. Forest loss, by contrast, reflects pressure across a landscape rather than a line on a map. Lenders with deforestation policies price forest loss across much wider areas, up to 50~km from a project.

The third element concerns time. Lenders with deforestation policies price the forest loss that a landscape has already experienced, but only weakly anticipate forest loss after the loan is signed. This is despite evidence that infrastructure projects contribute to forest loss in surrounding areas~\cite{sonter_mining_2017}. In terms of double materiality, this suggests that the pricing of project financing reflects nature-related risks borne by lenders more than the impacts of the projects themselves on nature. Loan spreads are set when the loan agreement is signed, so they cannot respond to subsequent damage caused by the project. Closing this gap would require loan terms that respond to environmental outcomes over the life of the loan. Satellite data now make it possible to periodically monitor forest loss around geolocated assets within their coverage~\cite{hansen_high-resolution_2013}. In the Brazilian Amazon, conditioning rural credit on environmental compliance reduced deforestation~\cite{assuncao_effect_2020}. Covenants or interest rate adjustments linked to verified land-cover changes could connect financing terms to outcomes over the life of a loan, provided that monitoring distinguishes the project's impact from wider changes in the landscape.

Pricing near protected areas appears stronger after the GBF, both for syndicates whose bookrunners hold ecosystem-related or species-related policies and for those with more Equator Principles signatories. This mirrors the evidence from the climate finance literature, which shows that banks began to price climate policy exposure and to reward green borrowers after the Paris Agreement~\cite{delis_being_2024,degryse_when_2023}. Stocks with large biodiversity footprints were similarly repriced around the Kunming Declaration~\cite{garel_investors_2024}. Global agreements may thus act as focal points that turn voluntary commitments into pricing  considerations. More stringent bank loan pricing is also higher near protected areas with stricter management categories, pointing towards the practice of lenders treating some designations as more binding than others. This is consistent with evidence showing that enforcement of nature protection has been priced in municipal bonds \cite{chen_pricing_2024}, and that new designations have reshaped the operations of nearby firms~\cite{akbari_real_2026}.

Shareholders also respond to proximity to protected areas. Sponsors experience short-term drops in stock prices when loans are signed for projects within 30~km of a protected area. While this is a much wider area than what lenders price, the price reactions do not establish a persistent valuation effect, as they are short-lived and fade within four days of signing the loan. We also find similar short-term drops in share prices for projects in landscapes that experience high subsequent forest loss. One explanation for the wider reach around protected areas lies in the architecture of project finance. Lenders protect themselves through the nexus of contracts that surrounds a project, including construction, supply and offtake agreements, sponsor support and covenants. These contracts can shift risk towards sponsors and other counterparties~\cite{corielli_risk_2010}. Hence, the cost of delays and overruns, and the reputational damage linked to nature, may be borne by sponsors' shareholders, who also have less detailed information about individual projects than arranging banks. 

\subsection*{Implications for financial policy}
For policymakers implementing Targets 14 and 15 of the GBF, our findings suggest that global commitments can be reflected in the price of capital for nature-exposed assets, and that lenders' own nature commitments are central to this transmission. However, many infrastructure financiers lack such commitments: only 17 of 65 infrastructure financiers of the Belt and Road Initiative require mitigation of biodiversity impact~\cite{narain_best-practice_2020}, and only 42\% of development banks that finance high-biodiversity-footprint infrastructure have biodiversity safeguards~\cite{narain_global_2023}. If projects that face higher prices from committed banks can turn to uncommitted ones, pricing may redistribute rather than reduce nature risk. This unintended effect has already been observed in fossil fuel finance, where banks in countries with strong divestment commitments and stringent environmental policies increased their financing of oil and gas companies abroad~\cite{cojoianu_does_2021}. For supervisors, geolocated project finance books offer a tractable window into banks' exposure to nature.

More broadly, nature should not be reduced to a single accounting unit for its risks to inform financing decisions. When capital is tied to a place, as it is in project finance, lenders and investors can, and some do, price nature risk. Whether they do so, however, appears to depend on the commitments of lenders and on the policy context. If the GBF is to align financial flows with nature, pricing will need to move beyond the boundaries of protected areas and beyond the moment a loan is signed, from where assets are located to how they impact nature over their lifetime.

\subsection*{Limitations and future research}
Our findings come with some limitations. First, our estimates are statistical relationships rather than causal effects: protected areas are not randomly located and lender policies are not randomly assigned. Second, because we observe only projects that received financing, we cannot tell whether banks with nature policies also deny credit to the most exposed projects, or whether those projects are simply financed by other lenders. Third, our measures of nature exposure are still coarse. We observe the coordinates of each project, but not how far it physically extends beyond the recorded coordinates. This risks understating the ecological footprint of infrastructure financing. Furthermore, forest loss around a project cannot be fully attributed to the project itself, particularly for a large buffer zone radii. Fourth, our policy measures record whether a bank has a relevant policy, not its stringency or implementation. Finally, our post-GBF results are tentative, given the short period since its adoption and the other initiatives and tighter monetary conditions that coincided with it. Future research should examine whether lenders with nature policies decline to finance exposed projects, whether such projects obtain financing elsewhere, and how loan terms respond to environmental outcomes after signing. 
\newpage
\section*{Methods}\label{sec:methods}

\subsection*{Project--loan data}
We identify project finance transactions in the LSEG SDC Platinum database~\cite{LSEG_SDC_2025}. We exclude projects that were not initiated and projects that are not classical project finance, following ref.~\cite{byoun_risk_2013}. Deal-level bank loan data are retrieved and matched from DealScan. Because a single project may be financed by several loan deals and a single loan deal may finance several projects, our unit of analysis is the project--loan pair. Our outcome of interest is the loan spread, defined as the interest and fees paid in basis points over the London Interbank Offered Rate (LIBOR) or an equivalent benchmark. The dependent variable is its natural logarithm, $\ln(\mathrm{Spread})$. 

The sample consists of 9,037 project--loan pairs for 3,793 projects in 115 countries, signed between 1990 and 2024 (Fig.~\ref{fig:map}). Variable definitions and summary statistics are reported in Supplementary Tables~\ref{revtab:definitions} and~\ref{revtab:summary_statistics}, and the industry composition of the sample in Supplementary Table~\ref{revtab:industry}. Estimation samples differ across models, depending on data availability: 6,907 project--loan pairs for models with lender policy ratios, 9,037 for models with the policy indicator, 3,720 for models with protected-area management categories, and 721--728 and 4,677 for models with forest loss before and after signing, respectively.

\subsection*{Lender nature policies}
We measure lenders' nature commitments at the level of the bookrunners (i.e., the banks that coordinate the loan). There are three categories of bank policies relevant to nature protection in our data: ecosystem-related, species-related and deforestation-related policies, which we collect for the bookrunners of each loan from BankTrack~\cite{banktrack_2025}. We also record whether each bookrunner is a signatory to the Equator Principles~\cite{equator_principles_2020}. 

For each loan, we construct three ratios that measure the presence and prevalence of nature-related policies within the arranging syndicate. The EcoSpecies ratio is the number of bookrunners with ecosystem-related or species-related policies divided by the total number of bookrunners in the loan deal. The Deforestation ratio is the number of bookrunners with deforestation-related policies divided by the total number of bookrunners. The EPFI ratio is the number of bookrunners that are Equator Principles signatories divided by the total number of bookrunners. Each ratio therefore ranges from 0 to 1. As an alternative measure, we use a binary EcoSpecies policy indicator reported in Supplementary Table~\ref{revtab:EcoSpecies_binary}.

\subsection*{Protected areas}
We measure a project's proximity to protected areas as the distance between the project's coordinates and the nearest protected area. We first geolocate each project using the point coordinates reported in SDC. Where coordinates are missing, we hand-collect them from Google Maps where the project site is identifiable (Fig.~\ref{fig:map}). We then compute the distance from each project to the nearest protected area in the World Database on Protected Areas (WDPA)~\cite{unepwcmc_wdpa_2024}. The WDPA, a joint product of the UN Environment Programme and the International Union for Conservation of Nature (IUCN), is the most comprehensive global database of marine and terrestrial protected areas. It records their boundaries, designation dates and management categories and contained 275,353 sites by 2023. We consider only protected areas designated before the date on which the loan was signed. Projects on land are matched to terrestrial or coastal protected areas, and marine projects to marine or coastal protected areas. From these distances, we construct indicators for projects located inside a protected area (0~m) or within 200~m, 400~m, 600~m, 800~m, 1~km, 1.5~km, 2~km, 3~km, 4~km and 5~km of one, and include each indicator in a separate regression. These thresholds define overlapping exposure groups rather than mutually exclusive distance bands.

We also use the IUCN management category of the nearest protected area~\cite{dudley_guidelines_2008} to measure the stringency of nature protection. These categories rank protected areas by how much human activity is allowed, ranging from I (strictly protected areas) to VI (protected areas allowing the sustainable use of natural resources). For ease of interpretation, we reverse the ordering so that the resulting score ranges from 1 to 6, with higher values indicating stricter protection. Project--loan pairs whose nearest protected area has no reported IUCN category are excluded from these models. 

\subsection*{Forest loss}
We measure forest loss around a project as the area of non-burn forest lost within a given distance of the project, before and after the loan is signed. Forest loss data is retrieved from the Global Forest Change dataset~\cite{hansen_high-resolution_2013}, which maps annual stand-replacing forest loss at 30~m resolution for 2001--2023. For each project, we calculate the area of forest lost in each year, in hectares, within circular buffers of 5~km to 50~km around the project. Because a substantial share of global forest loss is caused by wildfire rather than by permanent land-use change~\cite{curtis_classifying_2018}, we estimate the burned component using the Moderate Resolution Imaging Spectroradiometer (MODIS) MCD64A1 burned-area product~\cite{MODIS_Burned_2021,giglio_collection_2018} and subtract it from total forest loss to obtain annual non-burn forest loss. 

We then compute total non-burn forest loss for each project-loan pair over up to ten years before the loan is signed and up to ten years after it. Forest loss before signing measures the state of the landscape that lenders could observe when the loan was priced. Forest loss after signing is an ex post measure that was not available to lenders (nor equity market investors) at that time, and measures how the landscape changed once the project was financed. We transform cumulative loss in each period as $\ln(1+F)$, where $F$ is the cumulative non-burn forest-loss measure in the relevant buffer and period. The addition of one retains observations with zero measured loss.

\subsection*{Loan pricing models}
The baseline specification can be written as:
\begin{equation}
\ln(\mathrm{Spread}_{pl}) = \alpha + \beta E_{pl} + \gamma P_{l} + \delta\,(E_{pl}\times P_{l}) + \boldsymbol{\theta}^{\prime}\mathbf{X}_{pl} + \mu_{t} + \mu_{i} + \mu_{c} + \mu_{k} + \varepsilon_{pl},
\label{eq:pricing}
\end{equation}
where $p$ indexes projects and $l$ loans; $E_{pl}$ is a measure of nature exposure (a protected-area proximity indicator, or $\ln(1+F_{pl})$ for cumulative non-burn forest loss before or after signing); $P_{l}$ is the EcoSpecies, Deforestation or EPFI ratio of the loan, coded on a 0--1 scale; and $\mathbf{X}_{pl}$ contains the log loan amount, the log maturity in months, an indicator for performance pricing, the log total project cost, the number of sponsors, and indicators for government support, multilateral agency loans or grants, offtake agreements to purchase all or a substantial part of the project's output, and projects located in more than one country. $\mu_{t}$, $\mu_{i}$, $\mu_{c}$ and $\mu_{k}$ are fixed effects for the year of signing, project industry, project country and project type. Standard errors are clustered at the project level. *, ** and *** denote $p<0.10$, $p<0.05$ and $p<0.01$, respectively.

Our key coefficient of interest, $\delta$, measures how the association between nature exposure and loan spreads varies with lenders' nature-related policy coverage ($P_{l}$). Because the dependent variable is in logs, the coefficients are interpreted as percentage changes in loan spreads. For proximity to protected areas, increasing policy coverage by $\Delta P$ changes the spread on nearby projects, relative to more distant ones, by $100[\exp(\delta\Delta P)-1]\%$. For forest loss, $\beta+\delta P$ is approximately the elasticity of the spread with respect to forest loss, and $\delta$ measures how this elasticity changes with policy coverage.

We also test whether pricing changed after the adoption of the Kunming--Montreal Global Biodiversity Framework (GBF). The framework was adopted on 19 December 2022, during the second part of the fifteenth Conference of the Parties to the Convention on Biological Diversity~\cite{cbd_kunming-montreal_2022}. We interact the term $E_{pl}\times P_{l}$ in equation~\eqref{eq:pricing} with a post-GBF indicator. The indicator equals one for loans signed after that date. The coefficient on this triple interaction captures whether lenders with nature-related policies price nature exposure differently after the GBF.

We further test whether pricing varies with the stringency of protection. We interact the term $E_{pl}\times P_{l}$ in equation~\eqref{eq:pricing} with the management-category score of the nearest protected area. The coefficient on this triple interaction captures whether these lenders price proximity differently depending on how strictly the nearest protected area is protected.

\subsection*{Equity market response}

\begin{figure}[!htbp]
\centering
\includegraphics[width=0.96\linewidth]{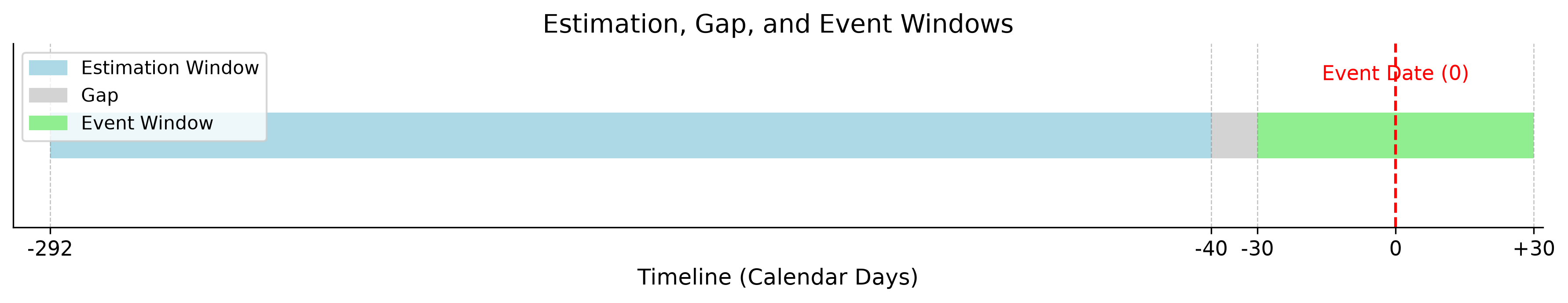}
\caption{\textbf{Event-study estimation and observation windows.} We estimate the five-factor model over calendar days $-292$ to $-41$. The event window runs from day $-30$ to day $+30$ around the loan signing date (day 0). We leave a ten-day gap (days $-40$ to $-31$) between the estimation window and the event window. Fig.~\ref{fig:event_study_combined} shows days $-5$ to $+5$.}
\label{fig:event_study_timeline}
\end{figure}

We restrict the sample of project-loan pairs to deals sponsored by listed firms headquartered in developed markets and use an event study~\cite{mackinlay_event_1997} to examine the stock price reaction to loan signings. The event date is the loan signing date. If a loan is signed on a Saturday or Sunday, we move the event date to the following Monday, and we treat loans signed by the same sponsor on the same day as one event. We measure event time in calendar days. For each firm and signing date, the event window covers 61 days, from 30 days before to 30 days after the signing. The estimation window runs from 292 to 41 days before the signing, so there is a 10-day gap between the estimation and event windows (Fig.~\ref{fig:event_study_timeline}). We drop events with fewer than seven returns in the estimation window. We estimate expected returns using the five-factor model of Fama and French~\cite{fama_five-factor_2015}. Specifically, we regress each firm's excess returns on the developed market excess return and the size (SMB), value (HML), profitability (RMW) and investment (CMA) factors, which we obtain from Kenneth R. French's data library~\cite{french_data_library}. Firm-level stock returns are from LSEG Workspace. The final sample consists of 1,900 project-loan signings by 601 firms between 1992 and 2024.

Using the estimated coefficients, we compute expected returns for each firm over the event window as the risk-free rate plus the estimated factor loadings multiplied by the factor returns; the intercept is not included. Abnormal returns are the difference between realised and expected returns. Although abnormal returns are computed over the full event window, the main results focus on days $-5$ to $+5$ (Fig.~\ref{fig:event_study_combined}). Specifically, for each signing, the cumulative abnormal return on day $t$ sums daily abnormal returns from day $-15$ to day $t$, $\mathrm{CAR}_i(t)=\sum_{\tau=-15}^{t}\mathrm{AR}_{i\tau}$, and we average these across signings to obtain $\mathrm{CAAR}(t)=N_t^{-1}\sum_{i=1}^{N_t}\mathrm{CAR}_i(t)$ for $t=-5,\dots,+5$, where $N_t$ is the number of signings with a stock return on day $t$. We report CAARs with 95\% confidence intervals based on cross-sectional standard errors. We compute CAARs for the full sample of signings and for several subsamples. First, we consider projects located within 10, 20, 30 and 40~km of a protected area. Second, we consider projects with non-burn forest loss within 10~km in the ten years after signing above the median and above the 75th percentile. Third, we consider projects within 10~km of a protected area whose forest loss is above the 75th percentile among these projects. We compute the percentiles across project--loan pairs. When a sponsor signs several loans on the same day, the event is included in a subsample if at least one of these loans meets the subsample's criterion, for example, if at least one project lies within 10 km of a protected area.

\section*{Acknowledgements \& Funding}
The authors are grateful for the comments of the participants to the JRC - EU Commission Conference on Nature Finance. Dr. Cojoianu acknowledges the kind support for the project from the EU Horizon and UKRI funded project Circhive (Grant agreement ID: 101082081) as well as funding from the SMU Singapore Green Finance Centre. Prof. Hoepner further acknowledges funding from the EU Horizon project  Nature-3B (Grant agreement ID: 101182455). Dr. Faour further acknowledges funding from the American University of Beirut University Research Board (Grant number 8820). 

\section*{Data availability}
The study uses project and financing records from SDC Platinum, sponsor return data from LSEG Workspace, lender policy records attributed to BankTrack, protected-area data from the WDPA, GLAD Global Forest Change data, MODIS MCD64A1 burned-area data and factors from the Kenneth R. French Data Library. Access to commercial financial records is subject to the respective providers' licensing conditions.

\bibliography{bibliography_combined}

\clearpage

\section*{Supplementary Information}
\begingroup
\setcounter{table}{0}
\captionsetup[table]{name=Supplementary Table}
\ManuscriptPrepareTables
\setlength{\parindent}{0pt}
\setlength{\ManuscriptTableMaxHeight}{0.86\textheight}
\begingroup
\renewcommand{\RevisedTableAlias}{revtab:definitions}
\begin{table}
\centering
\caption{Definitions of variables used in project-loan pair analysis.}\label{tabl:definitions}
\fontsize{9}{10.8}\selectfont
\setlength{\tabcolsep}{3.5pt}
\renewcommand{\arraystretch}{1.08}
\begin{tabularx}{\linewidth}{@{}>{\raggedright\arraybackslash}p{27mm}>{\raggedright\arraybackslash}X>{\raggedright\arraybackslash}p{24mm}@{}}
\toprule
Variable & \textbf{Definition} & \textbf{Source} \\
\midrule
Project Cost & Total project costs (\$millions) & SDC Project Finance \\
Project Size & The logarithm of the total project cost as a measure of project size &  \\
Spon & The number of sponsors in a project company &  \\
Gsup & The proportion (\%) of projects with government grants and support &  \\
Multia & A dummy variable indicating whether project finance involves multilateral agency loans or grants &  \\
Mult & A dummy variable indicating whether a project is located in more than 1 nation &  \\
Offt & A dummy variable indicating the existence of an offtake arrangement to purchase all or a substantial part of the product produced by the project &  \\
WDPA Distance (m) & The distance between a project location and its nearest WDPA-protected area (see Data and Methodology for more details). & Self-calculated \\
WDPA *m & A dummy variable indicating whether the WDPA distance is lesser than * m &  \\
WDPA stringency & A reversed version of the IUCN categories: it ranges from 1 to 6, with higher values indicating stricter levels of protection & IUCN’s WDPA \\
Kunming & A dummy variable indicating whether a loan deal was signed after the Kunming–Montreal Agreement (on 19 December 2022). &  \\
Forest Loss*km Pre & The cumulative non-burning forest loss within *km during the ten years preceding project financing, with the units measured in hectares & \cite{hansen_high-resolution_2013} \\
Forest Loss*km Post & The cumulative non-burning forest loss within *km during the ten years following project financing, with the units measured in hectares &  \\
KBA Distance (m) & The distance between a project location and its nearest KBA area & Self-calculated \\
KBA *m & A dummy variable indicating whether the KBA distance is greater than * m &  \\
Spread & The rate of interest and fees paid in basis points (bps) in excess of LIBOR or its equivalent & SDC \\
Ln Spread & The natural logarithm of Spread &  \\
Ln Loan Size & The natural logarithm of loan amount &  \\
Ln Maturity & The natural logarithm of the number of months to maturity &  \\
Performance Price & A dummy variable whether the loan has performance pricing provision &  \\
EcoSpecies Ratio & The number of bookrunners that have ecosystem related or species related policies divided by the total number of bookrunners in a loan deal. In alternative models, we also use the dummy eco-species indicator, which quantifies whether at least one bookrunner has an ecosystem-related or species policy. & BankTrack \\
Deforestation Ratio & The number of bookrunners with deforestation related policies divided by the total number of bookrunners in a loan deal. In alternative models, we also use the dummy deforestation policies indicator, which quantifies whether at least one bookrunner has a deforestation related policy.  &  \\
EPFIs Ratio & The number of bookrunners who are signatories to the Equator Principles divided by the total number of bookrunners in a loan deal. In alternative models, we also use the dummy EPFI indicator, which quantifies whether at least one bookrunner is an EPFI signatory.  &  \\
\bottomrule
\end{tabularx}
\end{table}

\endgroup
\clearpage
\setlength{\ManuscriptTableMaxHeight}{\dimexpr\textheight-2\baselineskip\relax}

\begingroup
\renewcommand{\RevisedTableAlias}{revtab:summary_statistics}
\begin{table}
\centering
\caption{Summary statistics.}\label{tab:summary_stats}
\fontsize{9.5}{11.5}\selectfont
\setlength{\tabcolsep}{3.5pt}
\renewcommand{\arraystretch}{1.06}
\begin{tabularx}{\linewidth}{@{}>{\raggedright\arraybackslash}X*{6}{r}@{}}
\toprule
 & N & Mean & SD & p25 & Median & p75 \\
\midrule
WDPA Distance (m) & 9037 & 179499 & 292651.4 & 7421 & 40513 & 226974 \\
WDPA stringency & 4645 & 2.81 & 1.36 & 2 & 3 & 3 \\
EcoSpecies Ratio & 6909 & .03 & .14 & 0 & 0 & 0 \\
Deforestation Ratio & 6909 & .02 & .11 & 0 & 0 & 0 \\
EPFIs Ratio & 6909 & .19 & .31 & 0 & 0 & .33 \\
KBA Distance (m) & 7299 & 722312.8 & 1380000 & 19442.48 & 184706.5 & 689443 \\
Ln Spread & 9037 & 5.17 & .77 & 4.79 & 5.16 & 5.66 \\
Ln Loan Size & 9037 & 4.62 & 1.62 & 3.58 & 4.74 & 5.8 \\
Ln Maturity & 9037 & 2.03 & .82 & 1.61 & 1.95 & 2.71 \\
Performance Price & 9037 & .99 & .09 & 1 & 1 & 1 \\
Project Size & 9037 & 6.21 & 1.45 & 5.3 & 6.21 & 7.15 \\
Spon & 9037 & 2.42 & 1.95 & 1 & 2 & 3 \\
Gsup & 9037 & .35 & .48 & 0 & 0 & 1 \\
Multia & 9037 & .08 & .27 & 0 & 0 & 0 \\
Mult & 9037 & .01 & .12 & 0 & 0 & 0 \\
Offt & 9037 & .26 & .44 & 0 & 0 & 1 \\
Kunming & 9037 & .08 & .27 & 0 & 0 & 0 \\
Forest Loss*km Post & 5870 & 3.35 & 2.97 & 0 & 3.85 & 6.02 \\
Forest Loss*km Pre & 1238 & 3.14 & 3.01 & 0 & 3.13 & 5.94 \\
\bottomrule
\end{tabularx}
\end{table}

\endgroup
\clearpage

\begingroup
\renewcommand{\RevisedTableAlias}{revtab:industry}
\begin{table}
\centering
\caption{Breakdown of project-loan pairs by industry.}\label{tab:project-loan_pairs}
\fontsize{9.5}{11.5}\selectfont
\setlength{\tabcolsep}{3.5pt}
\renewcommand{\arraystretch}{1.06}
\begin{tabularx}{\linewidth}{@{}>{\raggedright\arraybackslash}X*{6}{r}@{}}
\toprule
\textbf{Project Industry} & \textbf{Total} & \textbf{wdpa\_0} & \textbf{wdpa\_200} & \textbf{wdpa\_500} & \textbf{wdpa\_1k} & \textbf{wdpa\_2k} \\
\midrule
Air Transportation & 189 & 10 & 10 & 15 & 18 & 46 \\
Amusement \& Recreation & 163 & 1 & 3 & 3 & 3 & 9 \\
Cement \& Glass Products & 20 & 0 & 0 & 0 & 0 & 0 \\
Chemicals & 616 & 2 & 2 & 10 & 18 & 26 \\
Communication & 351 & 11 & 11 & 15 & 19 & 36 \\
Construction & 433 & 3 & 6 & 9 & 27 & 51 \\
Electronic And Electrical Products & 53 & 0 & 0 & 0 & 0 & 0 \\
Food & 13 & 0 & 0 & 0 & 0 & 0 \\
Healthcare & 153 & 0 & 0 & 0 & 8 & 8 \\
Highway Transportation & 689 & 8 & 30 & 36 & 47 & 67 \\
Lumber, Wood Products \& Paper & 33 & 3 & 3 & 4 & 4 & 4 \\
Machinery & 57 & 0 & 0 & 0 & 0 & 0 \\
Mining & 558 & 2 & 11 & 14 & 20 & 31 \\
Oil \& Gas & 2692 & 16 & 42 & 74 & 126 & 250 \\
Other Transportation Services & 257 & 4 & 4 & 4 & 27 & 34 \\
Others & 10 & 0 & 0 & 0 & 0 & 0 \\
Petroleum Refining, Rubber \& Plastics & 11 & 0 & 0 & 0 & 0 & 0 \\
Public Administration & 188 & 9 & 9 & 10 & 12 & 12 \\
Railroads & 48 & 0 & 0 & 4 & 4 & 11 \\
Steel \& Nonferrous Metals & 198 & 0 & 0 & 0 & 2 & 7 \\
Utilities & 2232 & 89 & 99 & 131 & 190 & 266 \\
Water Transportation & 73 & 0 & 0 & 0 & 3 & 4 \\
\bottomrule
\end{tabularx}
\end{table}

\endgroup
\clearpage

\begin{ManuscriptLandscapeTable}
\begin{table}
\centering
\caption{Model output for WDPA entry proximity different buffers and project-loan pair EcoSpecies ratio. Standard errors in parentheses with $^{*}p<0.10$, $^{**}p<0.05$ and $^{***}p<0.01$.}\label{tab:EcoSpecies_wdpa}
\fontsize{9}{11}\selectfont
\setlength{\tabcolsep}{2.8pt}
\renewcommand{\arraystretch}{1.0}
\begin{tabular*}{\linewidth}{@{\extracolsep{\fill}}>{\raggedright\arraybackslash}p{40mm}*{11}{r}@{}}
\toprule
 & (1) & (2) & (3) & (4) & (5) & (6) & (7) & (8) & (9) & (10) & (11) \\
WDPA threshold & Inside & 200 m & 400 m & 600 m & 800 m & 1 km & 1.5 km & 2 km & 3 km & 4 km & 5 km \\
\midrule
\multicolumn{12}{@{}l}{\textit{Dependent variable: ln\_spread}} \\
\addlinespace[3pt]
EcoSpecies\_ratio & 0.296*** & 0.274*** & 0.274*** & 0.276*** & 0.275*** & 0.274*** & 0.269*** & 0.285*** & 0.321*** & 0.319*** & 0.324*** \\
 & (0.073) & (0.070) & (0.071) & (0.071) & (0.071) & (0.071) & (0.072) & (0.074) & (0.074) & (0.073) & (0.073) \\
WDPA proximity & -0.146 & -0.058 & -0.027 & -0.010 & 0.011 & -0.005 & 0.017 & 0.026 & 0.036 & -0.008 & -0.004 \\
 & (0.144) & (0.106) & (0.077) & (0.070) & (0.061) & (0.053) & (0.045) & (0.043) & (0.037) & (0.038) & (0.037) \\
WDPA proximity $\times$ policy & 0.521 & 0.637*** & 0.531** & 0.489** & 0.436** & 0.427** & 0.423** & 0.178 & -0.048 & -0.029 & -0.043 \\
 & (0.320) & (0.236) & (0.229) & (0.224) & (0.206) & (0.196) & (0.175) & (0.190) & (0.176) & (0.162) & (0.158) \\
\midrule
ln\_loansize & -0.004 & -0.003 & -0.003 & -0.003 & -0.003 & -0.003 & -0.003 & -0.003 & -0.003 & -0.004 & -0.003 \\
 & (0.006) & (0.006) & (0.006) & (0.006) & (0.006) & (0.006) & (0.006) & (0.007) & (0.007) & (0.007) & (0.007) \\
ln\_maturity & 0.078*** & 0.077*** & 0.077*** & 0.077*** & 0.077*** & 0.077*** & 0.077*** & 0.077*** & 0.078*** & 0.078*** & 0.078*** \\
 & (0.013) & (0.013) & (0.013) & (0.013) & (0.014) & (0.014) & (0.014) & (0.014) & (0.013) & (0.013) & (0.013) \\
performance\_price & 0.255 & 0.261 & 0.261 & 0.261 & 0.260 & 0.259 & 0.259 & 0.256 & 0.255 & 0.258 & 0.257 \\
 & (0.165) & (0.164) & (0.164) & (0.164) & (0.164) & (0.164) & (0.164) & (0.164) & (0.164) & (0.164) & (0.164) \\
size & -0.053*** & -0.054*** & -0.054*** & -0.054*** & -0.054*** & -0.054*** & -0.055*** & -0.054*** & -0.054*** & -0.054*** & -0.054*** \\
 & (0.010) & (0.010) & (0.010) & (0.010) & (0.010) & (0.010) & (0.010) & (0.010) & (0.010) & (0.010) & (0.010) \\
spon & -0.000 & -0.000 & -0.000 & -0.000 & -0.000 & -0.000 & -0.001 & -0.001 & -0.001 & -0.001 & -0.001 \\
 & (0.007) & (0.007) & (0.007) & (0.007) & (0.007) & (0.007) & (0.007) & (0.007) & (0.007) & (0.007) & (0.007) \\
gsup & -0.048 & -0.049 & -0.048 & -0.048 & -0.048 & -0.048 & -0.049 & -0.048 & -0.047 & -0.047 & -0.047 \\
 & (0.035) & (0.035) & (0.035) & (0.035) & (0.035) & (0.035) & (0.035) & (0.035) & (0.035) & (0.035) & (0.035) \\
multia & 0.120*** & 0.120*** & 0.120*** & 0.121*** & 0.121*** & 0.122*** & 0.122*** & 0.121*** & 0.121*** & 0.121*** & 0.121*** \\
 & (0.045) & (0.045) & (0.045) & (0.045) & (0.045) & (0.045) & (0.045) & (0.045) & (0.045) & (0.045) & (0.045) \\
mult & -0.027 & -0.025 & -0.025 & -0.024 & -0.024 & -0.022 & -0.022 & -0.023 & -0.026 & -0.025 & -0.024 \\
 & (0.069) & (0.070) & (0.070) & (0.070) & (0.070) & (0.070) & (0.070) & (0.070) & (0.070) & (0.069) & (0.069) \\
offt & -0.035 & -0.035 & -0.035 & -0.036 & -0.035 & -0.036 & -0.035 & -0.035 & -0.035 & -0.035 & -0.035 \\
 & (0.029) & (0.029) & (0.029) & (0.029) & (0.029) & (0.029) & (0.029) & (0.029) & (0.029) & (0.029) & (0.029) \\
\midrule
Observations & 6907 & 6907 & 6907 & 6907 & 6907 & 6907 & 6907 & 6907 & 6907 & 6907 & 6907 \\
Adjusted R-squared & 0.321 & 0.321 & 0.321 & 0.321 & 0.321 & 0.321 & 0.321 & 0.321 & 0.321 & 0.320 & 0.320 \\
\bottomrule
\end{tabular*}
\par\vspace{5pt}
{\footnotesize\raggedright WDPA proximity is the indicator for the threshold shown in each column; Inside corresponds to wdpa\_0=1. Policy denotes the policy variable in the first coefficient row. All specifications include separate fixed effects for project type, project industry, loan-signing year and project country. Standard errors are clustered at the project level. \par}
\end{table}

\end{ManuscriptLandscapeTable}
\clearpage

\begingroup
\renewcommand{\RevisedTableAlias}{revtab:EcoSpecies_binary}
\begin{ManuscriptLandscapeTable}
\begin{table}
\centering
\caption{Model output for WDPA entry proximity different buffers and project-loan pair EcoSpecies ratio dummy = 1. Standard errors in parentheses with $^{*}p<0.10$, $^{**}p<0.05$ and $^{***}p<0.01$.}\label{tab:EcoSpecies_wdpa_dummy}
\fontsize{9}{11}\selectfont
\setlength{\tabcolsep}{2.8pt}
\renewcommand{\arraystretch}{1.0}
\begin{tabular*}{\linewidth}{@{\extracolsep{\fill}}>{\raggedright\arraybackslash}p{40mm}*{11}{r}@{}}
\toprule
 & (1) & (2) & (3) & (4) & (5) & (6) & (7) & (8) & (9) & (10) & (11) \\
WDPA threshold & Inside & 200 m & 400 m & 600 m & 800 m & 1 km & 1.5 km & 2 km & 3 km & 4 km & 5 km \\
\midrule
\multicolumn{12}{@{}l}{\textit{Dependent variable: ln\_spread}} \\
\addlinespace[3pt]
EcoSpecies\_d=1 & 0.040 & 0.034 & 0.032 & 0.033 & 0.034 & 0.034 & 0.033 & 0.040 & 0.062 & 0.049 & 0.056 \\
 & (0.040) & (0.040) & (0.040) & (0.040) & (0.040) & (0.041) & (0.042) & (0.043) & (0.039) & (0.039) & (0.040) \\
WDPA proximity & -0.184 & -0.102 & -0.081 & -0.054 & -0.030 & -0.027 & -0.008 & 0.007 & 0.016 & -0.024 & -0.018 \\
 & (0.114) & (0.089) & (0.072) & (0.064) & (0.057) & (0.050) & (0.043) & (0.041) & (0.034) & (0.034) & (0.032) \\
WDPA proximity $\times$ policy & 0.429** & 0.409** & 0.352** & 0.318** & 0.298** & 0.216** & 0.172** & 0.082 & -0.037 & 0.020 & -0.008 \\
 & (0.167) & (0.164) & (0.148) & (0.138) & (0.132) & (0.110) & (0.087) & (0.081) & (0.103) & (0.090) & (0.085) \\
\midrule
ln\_loansize & -0.006 & -0.006 & -0.006 & -0.006 & -0.006 & -0.006 & -0.006 & -0.005 & -0.005 & -0.006 & -0.006 \\
 & (0.006) & (0.006) & (0.006) & (0.006) & (0.006) & (0.006) & (0.006) & (0.006) & (0.006) & (0.006) & (0.006) \\
ln\_maturity & 0.074*** & 0.074*** & 0.073*** & 0.073*** & 0.073*** & 0.073*** & 0.073*** & 0.073*** & 0.073*** & 0.073*** & 0.074*** \\
 & (0.012) & (0.012) & (0.012) & (0.012) & (0.012) & (0.012) & (0.012) & (0.012) & (0.012) & (0.012) & (0.012) \\
performance\_price & 0.173 & 0.162 & 0.162 & 0.161 & 0.164 & 0.162 & 0.162 & 0.161 & 0.160 & 0.164 & 0.163 \\
 & (0.113) & (0.113) & (0.113) & (0.114) & (0.113) & (0.113) & (0.113) & (0.113) & (0.113) & (0.113) & (0.113) \\
size & -0.050*** & -0.050*** & -0.051*** & -0.051*** & -0.051*** & -0.051*** & -0.052*** & -0.051*** & -0.051*** & -0.051*** & -0.051*** \\
 & (0.011) & (0.011) & (0.011) & (0.011) & (0.011) & (0.011) & (0.011) & (0.011) & (0.011) & (0.011) & (0.011) \\
spon & -0.004 & -0.004 & -0.004 & -0.004 & -0.004 & -0.004 & -0.004 & -0.004 & -0.004 & -0.004 & -0.004 \\
 & (0.006) & (0.006) & (0.006) & (0.006) & (0.006) & (0.006) & (0.006) & (0.006) & (0.006) & (0.006) & (0.006) \\
gsup & -0.045 & -0.047 & -0.046 & -0.047 & -0.047 & -0.047 & -0.047 & -0.046 & -0.046 & -0.046 & -0.046 \\
 & (0.031) & (0.031) & (0.031) & (0.031) & (0.031) & (0.031) & (0.031) & (0.031) & (0.031) & (0.031) & (0.031) \\
multia & 0.076** & 0.077** & 0.077** & 0.078** & 0.079** & 0.080** & 0.081** & 0.081** & 0.082** & 0.080** & 0.081** \\
 & (0.035) & (0.035) & (0.035) & (0.035) & (0.035) & (0.035) & (0.035) & (0.035) & (0.035) & (0.035) & (0.035) \\
mult & 0.011 & 0.013 & 0.012 & 0.015 & 0.014 & 0.020 & 0.019 & 0.017 & 0.014 & 0.017 & 0.017 \\
 & (0.055) & (0.055) & (0.055) & (0.056) & (0.056) & (0.056) & (0.056) & (0.055) & (0.055) & (0.055) & (0.055) \\
offt & -0.064** & -0.064** & -0.064** & -0.065** & -0.065** & -0.065** & -0.064** & -0.064** & -0.065** & -0.064** & -0.064** \\
 & (0.027) & (0.027) & (0.027) & (0.027) & (0.027) & (0.027) & (0.027) & (0.027) & (0.027) & (0.027) & (0.027) \\
\midrule
Observations & 9037 & 9037 & 9037 & 9037 & 9037 & 9037 & 9037 & 9037 & 9037 & 9037 & 9037 \\
Adjusted R-squared & 0.313 & 0.313 & 0.313 & 0.313 & 0.312 & 0.312 & 0.312 & 0.312 & 0.312 & 0.312 & 0.312 \\
\bottomrule
\end{tabular*}
\par\vspace{5pt}
{\footnotesize\raggedright WDPA proximity is the indicator for the threshold shown in each column; Inside corresponds to wdpa\_0=1. Policy denotes the policy variable in the first coefficient row. All specifications include separate fixed effects for project type, project industry, loan-signing year and project country. Standard errors are clustered at the project level. \par}
\end{table}

\end{ManuscriptLandscapeTable}
\endgroup
\clearpage

\begin{table}
\centering
\caption{Model output for WDPA entry proximity different buffers, project-loan pair EcoSpecies ratio and pre/post-Kunming-Montreal. Standard errors in parentheses with $^{*}p<0.10$, $^{**}p<0.05$ and $^{***}p<0.01$.} \label{tab:EcoSpecies_wdpa_kunming}
\fontsize{9.5}{11.5}\selectfont
\setlength{\tabcolsep}{3.5pt}
\renewcommand{\arraystretch}{1.06}
\begin{tabularx}{\linewidth}{@{}>{\raggedright\arraybackslash}X*{4}{r}@{}}
\toprule
 & (1) & (2) & (3) & (4) \\
 & ln\_spread & ln\_spread & ln\_spread & ln\_spread \\
\midrule
EcoSpecies\_ratio & 0.226*** & 0.173 & 0.192** & 0.175 \\
 & (0.085) & (0.175) & (0.080) & (0.174) \\
wdpa\_0=1 & -0.124 & 0.885*** &  &  \\
 & (0.144) & (0.337) &  &  \\
wdpa\_0=1 \# EcoSpecies\_ratio & 0.503* & -0.774* &  &  \\
 & (0.304) & (0.447) &  &  \\
wdpa\_0=1 \# Kunming=1 \# EcoSpecies\_ratio & 6.407*** &  &  &  \\
 & (0.817) &  &  &  \\
wdpa\_0=1 \# EcoSpecies\_ratio \# wdpa\_stringency &  & 0.465** &  &  \\
 &  & (0.205) &  &  \\
wdpa\_200=1 &  &  & -0.043 & 0.939*** \\
 &  &  & (0.105) & (0.344) \\
wdpa\_200=1 \# EcoSpecies\_ratio &  &  & 0.674*** & -0.839* \\
 &  &  & (0.239) & (0.451) \\
wdpa\_200=1 \# Kunming=1 \# EcoSpecies\_ratio &  &  & 6.197*** &  \\
 &  &  & (0.796) &  \\
wdpa\_200=1 \# EcoSpecies\_ratio \# wdpa\_stringency &  &  &  & 0.471** \\
 &  &  &  & (0.203) \\
\midrule
Observations & 6907 & 3720 & 6907 & 3720 \\
Adjusted R-squared & 0.321 & 0.277 & 0.322 & 0.277 \\
\bottomrule
\end{tabularx}
\par\vspace{5pt}
{\footnotesize\raggedright WDPA proximity is the indicator for the threshold shown in each row; Inside corresponds to wdpa\_0=1. All specifications include separate fixed effects for project type, project industry, loan-signing year and project country. Standard errors are clustered at the project level. \par}
\end{table}

\clearpage

\begingroup
\renewcommand{\RevisedTableAlias}{revtab:EPFIs_wdpa_kunming}
\begin{table}
\centering
\caption{Model output for WDPA entry proximity different buffers and project-loan pair EPFI ratio. Standard errors in parentheses with $^{*}p<0.10$, $^{**}p<0.05$ and $^{***}p<0.01$.}\label{tab:EPFI_wdpa}
\fontsize{9.5}{11.5}\selectfont
\setlength{\tabcolsep}{3.5pt}
\renewcommand{\arraystretch}{1.06}
\begin{tabularx}{\linewidth}{@{}>{\raggedright\arraybackslash}X*{4}{r}@{}}
\toprule
 & (1) & (2) & (3) & (4) \\
 & ln\_spread & ln\_spread & ln\_spread & ln\_spread \\
\midrule
EPFIs\_ratio & 0.015 & -0.005 & 0.012 & -0.005 \\
 & (0.040) & (0.104) & (0.040) & (0.104) \\
wdpa\_0=1 & -0.268 & 0.886*** &  &  \\
 & (0.214) & (0.306) &  &  \\
wdpa\_0=1 \# EPFIs\_ratio & 0.519 & -0.741* &  &  \\
 & (0.371) & (0.406) &  &  \\
wdpa\_0=1 \# Kunming=1 \# EPFIs\_ratio & 3.701*** &  &  &  \\
 & (0.585) &  &  &  \\
wdpa\_0=1 \# EPFIs\_ratio \# wdpa\_stringency &  & 0.455* &  &  \\
 &  & (0.241) &  &  \\
wdpa\_200=1 &  &  & -0.095 & 0.953*** \\
 &  &  & (0.145) & (0.333) \\
wdpa\_200=1 \# EPFIs\_ratio &  &  & 0.414 & -0.814* \\
 &  &  & (0.300) & (0.436) \\
wdpa\_200=1 \# Kunming=1 \# EPFIs\_ratio &  &  & 3.798*** &  \\
 &  &  & (0.544) &  \\
wdpa\_200=1 \# EPFIs\_ratio \# wdpa\_stringency &  &  &  & 0.468* \\
 &  &  &  & (0.239) \\
\midrule
Observations & 6907 & 3720 & 6907 & 3720 \\
Adjusted R-squared & 0.319 & 0.275 & 0.319 & 0.275 \\
\bottomrule
\end{tabularx}
\par\vspace{5pt}
{\footnotesize\raggedright WDPA proximity is the indicator for the threshold shown in each row; Inside corresponds to wdpa\_0=1. All specifications include separate fixed effects for project type, project industry, loan-signing year and project country. Standard errors are clustered at the project level. \par}
\end{table}

\endgroup
\clearpage

\begin{ManuscriptLandscapeTable}
\begin{table}
\centering
\caption{Model output for WDPA entry proximity different buffers and project-loan pair deforestation ratio. Standard errors in parentheses with $^{*}p<0.10$, $^{**}p<0.05$ and $^{***}p<0.01$.}\label{tab:deforestation_wdpa}
\fontsize{9}{11}\selectfont
\setlength{\tabcolsep}{2.8pt}
\renewcommand{\arraystretch}{1.0}
\begin{tabular*}{\linewidth}{@{\extracolsep{\fill}}>{\raggedright\arraybackslash}p{40mm}*{11}{r}@{}}
\toprule
 & (1) & (2) & (3) & (4) & (5) & (6) & (7) & (8) & (9) & (10) & (11) \\
WDPA threshold & Inside & 200 m & 400 m & 600 m & 800 m & 1 km & 1.5 km & 2 km & 3 km & 4 km & 5 km \\
\midrule
\multicolumn{12}{@{}l}{\textit{Dependent variable: ln\_spread}} \\
\addlinespace[3pt]
Deforestation\_ratio & 0.414*** & 0.383*** & 0.381*** & 0.382*** & 0.382*** & 0.383*** & 0.382*** & 0.400*** & 0.394*** & 0.397*** & 0.388*** \\
 & (0.086) & (0.082) & (0.082) & (0.082) & (0.083) & (0.084) & (0.084) & (0.083) & (0.088) & (0.084) & (0.084) \\
WDPA proximity & -0.143 & -0.048 & -0.025 & -0.007 & 0.012 & -0.003 & 0.022 & 0.029 & 0.030 & -0.013 & -0.011 \\
 & (0.143) & (0.105) & (0.077) & (0.070) & (0.061) & (0.052) & (0.045) & (0.042) & (0.038) & (0.038) & (0.036) \\
WDPA proximity $\times$ policy & 0.398 & 0.515** & 0.509** & 0.486** & 0.419** & 0.390** & 0.392** & 0.179 & 0.142 & 0.101 & 0.124 \\
 & (0.315) & (0.236) & (0.220) & (0.218) & (0.202) & (0.196) & (0.188) & (0.241) & (0.180) & (0.177) & (0.172) \\
\midrule
ln\_loansize & -0.003 & -0.003 & -0.003 & -0.003 & -0.003 & -0.003 & -0.003 & -0.003 & -0.003 & -0.003 & -0.003 \\
 & (0.006) & (0.006) & (0.006) & (0.006) & (0.006) & (0.006) & (0.006) & (0.007) & (0.007) & (0.007) & (0.007) \\
ln\_maturity & 0.077*** & 0.076*** & 0.076*** & 0.076*** & 0.076*** & 0.076*** & 0.076*** & 0.076*** & 0.076*** & 0.077*** & 0.077*** \\
 & (0.013) & (0.014) & (0.014) & (0.013) & (0.014) & (0.014) & (0.014) & (0.014) & (0.013) & (0.013) & (0.013) \\
performance\_price & 0.251 & 0.257 & 0.258 & 0.258 & 0.258 & 0.257 & 0.256 & 0.254 & 0.254 & 0.257 & 0.257 \\
 & (0.164) & (0.163) & (0.163) & (0.163) & (0.163) & (0.163) & (0.163) & (0.163) & (0.163) & (0.163) & (0.163) \\
size & -0.053*** & -0.053*** & -0.053*** & -0.054*** & -0.054*** & -0.054*** & -0.054*** & -0.054*** & -0.053*** & -0.053*** & -0.053*** \\
 & (0.010) & (0.010) & (0.010) & (0.010) & (0.010) & (0.010) & (0.010) & (0.010) & (0.010) & (0.010) & (0.010) \\
spon & -0.000 & -0.000 & -0.000 & -0.000 & -0.001 & -0.001 & -0.001 & -0.001 & -0.001 & -0.001 & -0.001 \\
 & (0.007) & (0.007) & (0.007) & (0.007) & (0.007) & (0.007) & (0.007) & (0.007) & (0.007) & (0.007) & (0.007) \\
gsup & -0.046 & -0.047 & -0.047 & -0.047 & -0.047 & -0.047 & -0.047 & -0.047 & -0.046 & -0.046 & -0.046 \\
 & (0.035) & (0.035) & (0.035) & (0.035) & (0.035) & (0.035) & (0.035) & (0.035) & (0.035) & (0.035) & (0.035) \\
multia & 0.123*** & 0.123*** & 0.122*** & 0.123*** & 0.123*** & 0.124*** & 0.124*** & 0.123*** & 0.124*** & 0.124*** & 0.124*** \\
 & (0.045) & (0.045) & (0.045) & (0.045) & (0.045) & (0.045) & (0.045) & (0.045) & (0.045) & (0.045) & (0.045) \\
mult & -0.021 & -0.020 & -0.020 & -0.020 & -0.019 & -0.017 & -0.018 & -0.019 & -0.019 & -0.018 & -0.019 \\
 & (0.070) & (0.070) & (0.070) & (0.070) & (0.070) & (0.070) & (0.070) & (0.070) & (0.070) & (0.070) & (0.070) \\
offt & -0.035 & -0.035 & -0.036 & -0.036 & -0.035 & -0.036 & -0.035 & -0.035 & -0.035 & -0.035 & -0.036 \\
 & (0.030) & (0.029) & (0.029) & (0.029) & (0.029) & (0.029) & (0.029) & (0.029) & (0.029) & (0.029) & (0.029) \\
\midrule
Observations & 6907 & 6907 & 6907 & 6907 & 6907 & 6907 & 6907 & 6907 & 6907 & 6907 & 6907 \\
Adjusted R-squared & 0.321 & 0.321 & 0.321 & 0.321 & 0.321 & 0.321 & 0.321 & 0.321 & 0.321 & 0.321 & 0.321 \\
\bottomrule
\end{tabular*}
\par\vspace{5pt}
{\footnotesize\raggedright WDPA proximity is the indicator for the threshold shown in each column; Inside corresponds to wdpa\_0=1. All specifications include separate fixed effects for project type, project industry, loan-signing year and project country. Standard errors are clustered at the project level. \par}
\end{table}

\end{ManuscriptLandscapeTable}
\clearpage

\begingroup
\renewcommand{\RevisedTableAlias}{revtab:prior_forest_loss}
\begin{ManuscriptLandscapeTable}
\begin{table}
\centering
\caption{Model output for prior non burn deforestation at different buffers and project-loan pair deforestation ratio. Standard errors in parentheses with $^{*}p<0.10$, $^{**}p<0.05$ and $^{***}p<0.01$.}\label{tab:non-burn_deforestation_pre}
\fontsize{9.5}{11.5}\selectfont
\setlength{\tabcolsep}{2.8pt}
\renewcommand{\arraystretch}{1.0}
\begin{tabular*}{\linewidth}{@{\extracolsep{\fill}}>{\raggedright\arraybackslash}p{58mm}*{6}{r}@{}}
\toprule
 & (1) & (2) & (3) & (4) & (5) & (6) \\
Buffer radius & 5 km & 10 km & 20 km & 30 km & 40 km & 50 km \\
\midrule
\multicolumn{7}{@{}l}{\textit{Dependent variable: ln\_spread}} \\
\addlinespace[3pt]
Prior forest loss (logged) & -0.008 & -0.008 & -0.014 & -0.010 & -0.008 & -0.008 \\
 & (0.011) & (0.009) & (0.009) & (0.008) & (0.008) & (0.008) \\
Policy $\times$ forest loss & 0.596*** & 0.473*** & 0.578*** & 0.537*** & 0.505*** & 0.496*** \\
 & (0.226) & (0.178) & (0.158) & (0.153) & (0.149) & (0.153) \\
Deforestation\_ratio & -2.466** & -2.539** & -4.189*** & -4.244*** & -4.267*** & -4.497*** \\
 & (1.200) & (1.223) & (1.276) & (1.355) & (1.420) & (1.556) \\
\midrule
ln\_loansize & 0.038*** & 0.039*** & 0.038*** & 0.037*** & 0.038*** & 0.037*** \\
 & (0.012) & (0.012) & (0.012) & (0.012) & (0.012) & (0.012) \\
ln\_maturity & 0.028 & 0.029 & 0.034 & 0.030 & 0.029 & 0.030 \\
 & (0.035) & (0.034) & (0.035) & (0.035) & (0.035) & (0.035) \\
performance\_price & 1.177*** & 1.164*** & 1.173*** & 1.165*** & 1.167*** & 1.171*** \\
 & (0.162) & (0.160) & (0.166) & (0.165) & (0.165) & (0.163) \\
size & -0.059*** & -0.060*** & -0.059*** & -0.060*** & -0.060*** & -0.059*** \\
 & (0.015) & (0.015) & (0.015) & (0.015) & (0.015) & (0.015) \\
spon & -0.020 & -0.020 & -0.019 & -0.020 & -0.020 & -0.020 \\
 & (0.015) & (0.015) & (0.014) & (0.014) & (0.014) & (0.014) \\
gsup & 0.002 & -0.000 & -0.009 & -0.004 & -0.001 & -0.004 \\
 & (0.069) & (0.068) & (0.070) & (0.069) & (0.069) & (0.068) \\
multia & 0.025 & 0.019 & 0.023 & 0.012 & 0.079 & 0.078 \\
 & (0.140) & (0.136) & (0.132) & (0.130) & (0.152) & (0.151) \\
mult & 0.000 & 0.000 & 0.000 & 0.000 & 0.000 & 0.000 \\
 & (.) & (.) & (.) & (.) & (.) & (.) \\
offt & 0.029 & 0.031 & 0.042 & 0.047 & 0.052 & 0.053 \\
 & (0.072) & (0.069) & (0.071) & (0.070) & (0.071) & (0.071) \\
\midrule
Observations & 723 & 728 & 723 & 726 & 721 & 724 \\
Adjusted R-squared & 0.284 & 0.284 & 0.288 & 0.286 & 0.286 & 0.288 \\
\bottomrule
\end{tabular*}
\par\vspace{5pt}
{\footnotesize\raggedright Each column uses the logged non-burning forest-loss measure for the buffer radius shown. Policy denotes Deforestation\_ratio. All specifications include separate fixed effects for project type, project industry, loan-signing year and project country. Standard errors are clustered at the project level.\par}
\end{table}

\end{ManuscriptLandscapeTable}
\endgroup
\clearpage

\begingroup
\renewcommand{\RevisedTableAlias}{revtab:subsequent_forest_loss}
\begin{ManuscriptLandscapeTable}
\begin{table}
\centering
\caption{Model output for subsequent non burn deforestation at different buffers and project-loan pair deforestation ratio. Standard errors in parentheses with $^{*}p<0.10$, $^{**}p<0.05$ and $^{***}p<0.01$.}\label{tab:non-burn_deforestation_post}
\fontsize{9.5}{11.5}\selectfont
\setlength{\tabcolsep}{2.8pt}
\renewcommand{\arraystretch}{1.0}
\begin{tabular*}{\linewidth}{@{\extracolsep{\fill}}>{\raggedright\arraybackslash}p{58mm}*{6}{r}@{}}
\toprule
 & (1) & (2) & (3) & (4) & (5) & (6) \\
Buffer radius & 5 km & 10 km & 20 km & 30 km & 40 km & 50 km \\
\midrule
\multicolumn{7}{@{}l}{\textit{Dependent variable: ln\_spread}} \\
\addlinespace[3pt]
Subsequent forest loss (logged) & -0.004 & -0.005 & -0.002 & -0.001 & -0.001 & 0.000 \\
 & (0.006) & (0.006) & (0.006) & (0.006) & (0.006) & (0.006) \\
Policy $\times$ forest loss & 0.156* & 0.151** & 0.147* & 0.121 & 0.126* & 0.133* \\
 & (0.086) & (0.071) & (0.083) & (0.079) & (0.073) & (0.069) \\
Deforestation\_ratio & -0.505 & -0.676 & -0.876 & -0.740 & -0.863 & -1.019 \\
 & (0.641) & (0.598) & (0.866) & (0.938) & (0.911) & (0.887) \\
\midrule
ln\_loansize & -0.003 & -0.003 & -0.003 & -0.003 & -0.003 & -0.003 \\
 & (0.008) & (0.008) & (0.008) & (0.008) & (0.008) & (0.008) \\
ln\_maturity & 0.081*** & 0.081*** & 0.081*** & 0.081*** & 0.081*** & 0.081*** \\
 & (0.017) & (0.017) & (0.017) & (0.017) & (0.017) & (0.017) \\
performance\_price & 0.472* & 0.471* & 0.474* & 0.474* & 0.474* & 0.474* \\
 & (0.252) & (0.252) & (0.253) & (0.253) & (0.253) & (0.253) \\
size & -0.057*** & -0.058*** & -0.057*** & -0.057*** & -0.057*** & -0.057*** \\
 & (0.013) & (0.013) & (0.013) & (0.013) & (0.013) & (0.013) \\
spon & 0.005 & 0.005 & 0.005 & 0.005 & 0.005 & 0.005 \\
 & (0.008) & (0.008) & (0.008) & (0.008) & (0.008) & (0.008) \\
gsup & -0.091* & -0.092* & -0.091* & -0.091* & -0.091* & -0.090* \\
 & (0.050) & (0.050) & (0.050) & (0.050) & (0.050) & (0.050) \\
multia & 0.126** & 0.125** & 0.126** & 0.126** & 0.126** & 0.126** \\
 & (0.060) & (0.060) & (0.060) & (0.060) & (0.060) & (0.060) \\
mult & -0.014 & -0.015 & -0.019 & -0.019 & -0.019 & -0.020 \\
 & (0.111) & (0.111) & (0.111) & (0.111) & (0.111) & (0.111) \\
offt & -0.056 & -0.057 & -0.056 & -0.056 & -0.056 & -0.056 \\
 & (0.037) & (0.037) & (0.037) & (0.037) & (0.037) & (0.037) \\
\midrule
Observations & 4677 & 4677 & 4677 & 4677 & 4677 & 4677 \\
Adjusted R-squared & 0.323 & 0.323 & 0.323 & 0.323 & 0.323 & 0.323 \\
\bottomrule
\end{tabular*}
\par\vspace{5pt}
{\footnotesize\raggedright Each column uses the logged non-burning forest-loss measure for the buffer radius shown. Policy denotes Deforestation\_ratio. All specifications include separate fixed effects for project type, project industry, loan-signing year and project country. Standard errors are clustered at the project level.\par}
\end{table}

\end{ManuscriptLandscapeTable}
\endgroup
\clearpage

\endgroup

\end{document}